\documentclass[twocolumn,prl]{revtex4}
\usepackage{graphicx}
\usepackage{amsmath}
\usepackage{amssymb}
\usepackage{wasysym}

\newcommand{\unit}[1]{\ensuremath{\, \mathrm{#1}}}

\begin{document}


\title{Evaporation-triggered Wetting Transition for Water Droplets upon Hydrophobic Microstructures}

\author{Peichun Tsai$^1$, Rob G. H. Lammertink$^2$, Matthias Wessling$^2$, and Detlef Lohse$^1$}

\affiliation{$^{1}$Physics of Fluids Group; $^{2}$Membrane Science and Technology Group, Faculty of Science and Technology, Impact and Mesa$^+$ Institutes, University of Twente, 7500AE Enschede, The Netherlands.}
\date{\today}

\begin{abstract}
When placed on rough hydrophobic surfaces, water droplets of diameter larger than a few millimeters can easily form pearls, as they are in the Cassie-Baxter state with air pockets trapped underneath the droplet. Intriguingly, a natural evaporating process can drive such a Fakir drop into a completely wetting (Wenzel) state. Our microscopic observations with simultaneous side and bottom views of evaporating droplets upon transparent hydrophobic microstructures elucidate the water-filling dynamics and the mechanism of this evaporation-triggered transition. For the present material the wetting transition occurs when the water droplet size decreases to a few hundreds of micrometers in radius. We present a general global energy argument which estimates the interfacial energies depending on the drop size and can account for the critical radius for the transition.  
\end{abstract}


\maketitle

In nature some plants with microstructured waxy leaves keep themselves dry and clean  against rain or pesticide droplets~\cite{waxy_leaves}.  
The hydrophobic, rough or even micro-structured surfaces suspend tiny liquid drops with air being trapped in between the droplets and the surface. This so-called Cassie-Baxter (CB) or ``Fakir'' state~\cite{CB_state} possesses a large contact angle and sometimes is highly disadvantageous in industrial applications, such as printing, coating, spraying, and sputtering processes. In contrast, in microfluidic applications the CB state is desirable to allow for slip and thus an enhanced flow rate~\cite{Lauga_JFM2003}.
In this Letter, we demonstrate that an evaporation process can drive the transition from a Fakir droplet to a homogeneously wetting state, i.e., to the Wenzel (W) state.  We explain the underlying mechanism by an energy balance argument.

The deceptively simple process of a freely evaporating droplet on a solid surface can produce various intriguing patterns, such as coffee stains~\cite{Deegan_Nature1997,Deegan_PRE2000} and wine tears~\cite{Hosoi_JFM_2001}. These patterns are consequences of the complex interplay between several physical processes: mass, heat and energy exchange across different interfaces, diffusive and convective flows, possible Marangoni circulations~\cite{marangoni} in the presence of a temperature or concentration gradient, and the movement of the triple line~\cite{review_evaporation_at_contact_line}. Yet, most investigations of evaporating droplets are conducted with hydrophilic and hydrophobic flat surfaces~\cite{Shin_2009,Erbil_Langmuir_1999,Erbil_Langmuir_2002,Cachile_Langmuir_2002,DBonn_JFM2006}, and rarely with ultra-hydrophobic patterned substrates~\cite{MacHale_Langmuir_2005,Zhang_2006,Jung_2008}.

\begin{figure}[b]
\begin{center}
\includegraphics[width=3.0in]{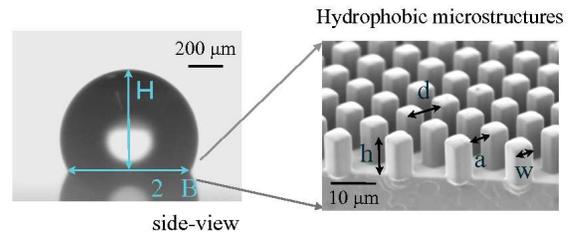}
\caption{\label{setup}(Color online) Experiment of an evaporating water droplet on a hydrophobic microstructured surface shown by the SEM image on the right.}
\end{center}
\end{figure}

Fig.~\ref{setup} shows the experiment of a freely evaporating water droplet placed on a hydrophobic micropatterned substrate. The surface is composed of PDMS (Polydimethylsiloxane) microstructures, thereby being transparent and ultra-hydrophobic. The precisely controllable microstructures were realized via a micro-molding technique. The sample preparation includes mixing and then degassing the PDMS component A with the curing agent B ($10:1$ mass ratio, Dowcorning Sylgard elastomer).  The mixture is cast onto a hydrophobized wafer with desired micropatterns and finally heated in an oven at $85^\circ$ for $3$ hours.
The SEM (scanning electron microscope) picture in Fig.~\ref{setup} shows the substrate, which consists of regular micron-sized pillars of width $w$, height $h$, and interspace $a$ in a square lattice with a periodicity $d = w + a$. In the experiments $w = 5~\unit{\mu m}$, $a = 5~\unit{\mu m}$, and $h$ is varied between $ 6,~10$, and $20~\unit{\mu m}$ to investigate the geometric effect on the wetting transition. 
Our micro-pillars are densely packed, with an area packing fraction of the solid pillars $\Phi_s = w^2/d^2 = 0.25$ ($\Phi_s = \pi w^2/4d^2 \simeq 0.20$) for rectangular (cylindrical) shaped pillars in a square lattice, where $\Phi_s$ is the ratio of the solid to the total cross-section areas.

The evaporation process was carried out at room temperature $(21 \pm 1)^\circ$C with a relative humidity of $35 \pm 5~\%$. The used liquid was ultra-purified milli-Q water with different initial drop sizes, $2-3~\unit{mm}$ in diameter. The side-view images of the evaporating drop were captured via $4 - 20\times$ magnifying lenses with a CCD camera using the recording rate of $1 - 2$ fps (frame per second). The temporal variations of drop height $H$ and base diameter $2B$ were measured.  Simultaneously, the sample was observed using an inverted microscope (Axiovert 40 CFL, Carl Zeiss BV). The bottom-view images were recorded through the translucent polymeric substrate with a high speed camera at $50 - 10~000$ fps via an objective of $10$ or $20~\times$ magnification. The highest spatial resolution is about $1~\unit{\mu m/pixel}$.  The snapshots of the bottom-views facilitate the investigation of the water infiltration dynamics and the accurate measurement of the critical base radius $B_c$ at the wetting transition.

\begin{figure}
\begin{center}
\includegraphics[width=3.2in]{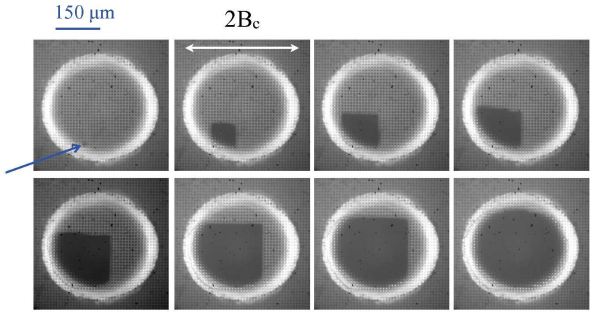}
\caption{\label{bottom-view} 
(Color online) Snapshots of the bottom-view reveal the water infiltration dynamics at the transition from a Cassie-Baxter to a Wenzel wetting state triggered by evaporation. Here, the square lattice have the dimension with $a\, =\, 5, w\, =\, 5,$ and $h\, =\, 10\,\unit{\mu m}$. The dark areas indicate the water imbibition, while the air pockets present in the bright areas enclosed by a rather bright circumference marked by the droplet base. These images are sequentially recorded at $t = 0,~30,~58,~72,~116,~180,~236$ and $310~\unit{ms}$ from the transitional point, determined by the first frame where a small initial infiltration point is observed (marked by an arrow).}
\end{center}
\end{figure}

Fig.~\ref{bottom-view} illustrates representative water impaling dynamics at the evaporation triggered CB to W wetting transition (1$^{st}$ frame) and beyond up to the fully wetting state (last frame).  In the Fakir state, prior to the transition we observe the perimeter of the contact line globally resembles a circular shape while locally adapting to the micro square lattice. During the evaporation the triple contact line advances in a stepwise manner, corresponding to the periodicity $d$ of the microstructures. We noticed that the shape of the droplet base at the critical point is mostly circular for tall micro-pillars ($h = 10$ and $20~\unit{\mu m}$). Polygonal or irregular base shapes were seen for short pillars ($h = 6~\unit{\mu m}$).  The infiltration point (see arrow in Fig.~\ref{bottom-view}) mostly ($\approx 93\%$ based on $28$ experiments) occurs at the edge of water base, i.e., close to the contact line. 

The water invading dynamics, as shown in Fig.~\ref{bottom-view}, reflects the geometric arrangement of the micro-pillars and presents two rather different time scales for the water propagation.  The wetted area is shown by the dark region with a square edge at the right upper corner, while the left bottom corner is constrained by the circular contact line. The water front first locally advances into one row and then quickly zips to the side perpendicular to the front direction. The water front propagated with a mean speed of $0.65~\unit{mm/s}$ while the fast zipping speed was about $8\times$ faster.  
The same kind of water impaling dynamics has been observed and more detailedly investigated with a variety of polymeric microstructures, though, at a spontaneous or a pre-triggered wetting transition~\cite{sbragaglia_PRL_2007,pirat_EPL_2008, peters_EPJE_2009}. Pronounced geometric effects on the water imbibition dynamics were presented for both hydrophobic~\cite{sbragaglia_PRL_2007} and hydrophilic micro-pillars~\cite{courbin_nature_mat_2007}.

\begin{figure}
\begin{center}
\includegraphics[width=3.0in]{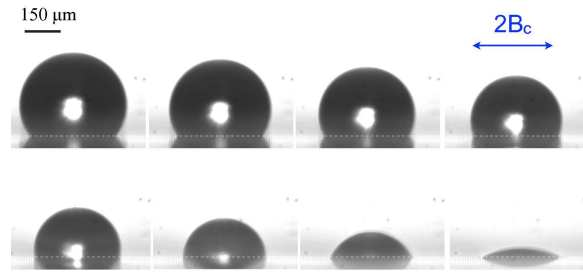}
\caption{\label{side-views}(Color online) Snapshots of the side-views of an evaporating droplet with the time interval $15$~s between the images, revealing the decrease of the contact angle.  The dash lines, marking the drop base, separate the main droplet from its mirror image. The arrow shows the length scale of the critical diameter $2B_c$ at the wetting transition, determined from the bottom views.
}
\end{center}
\end{figure}

\begin{figure}
\begin{center}
\includegraphics[width=2.8in]{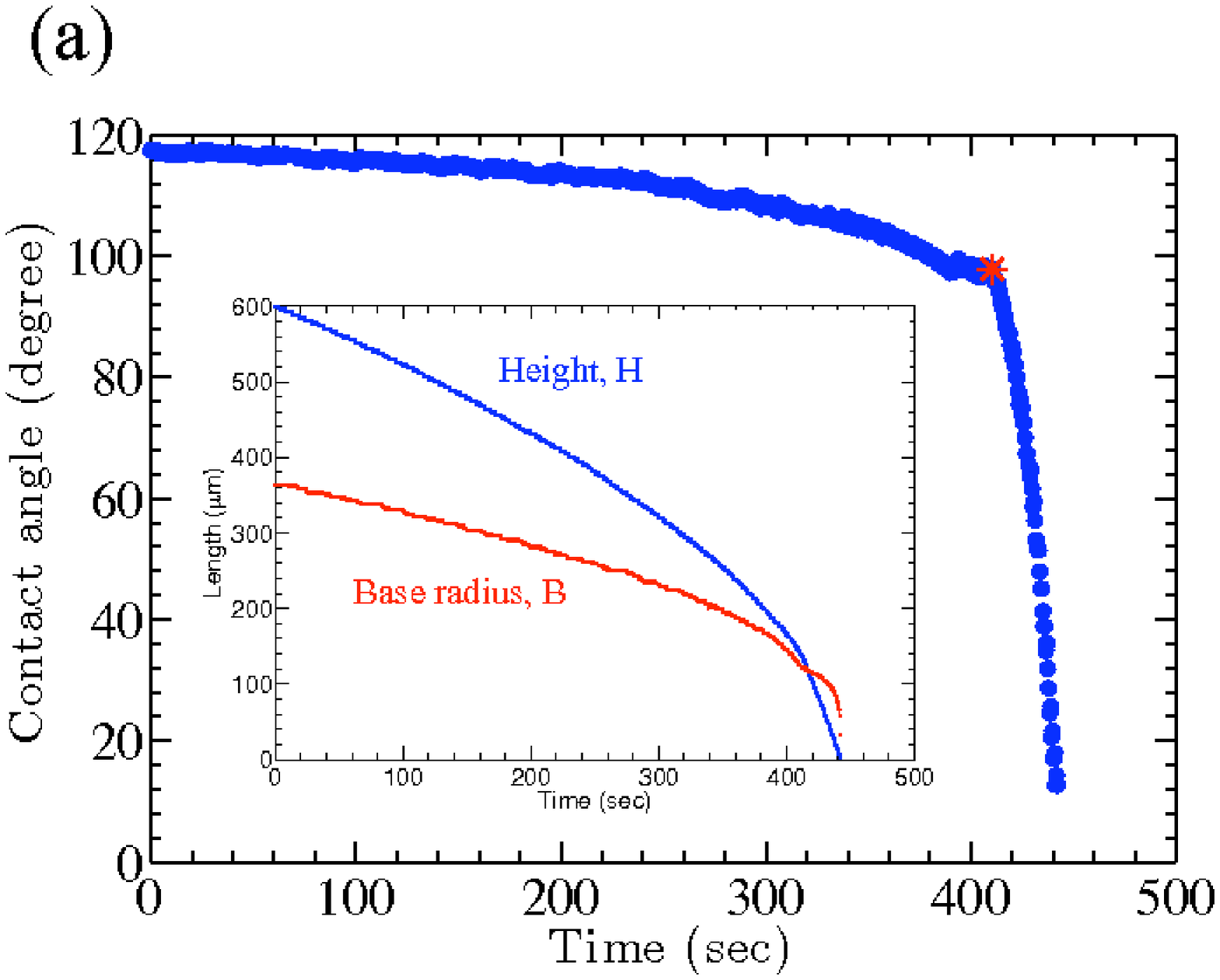}
\includegraphics[width=2.8in]{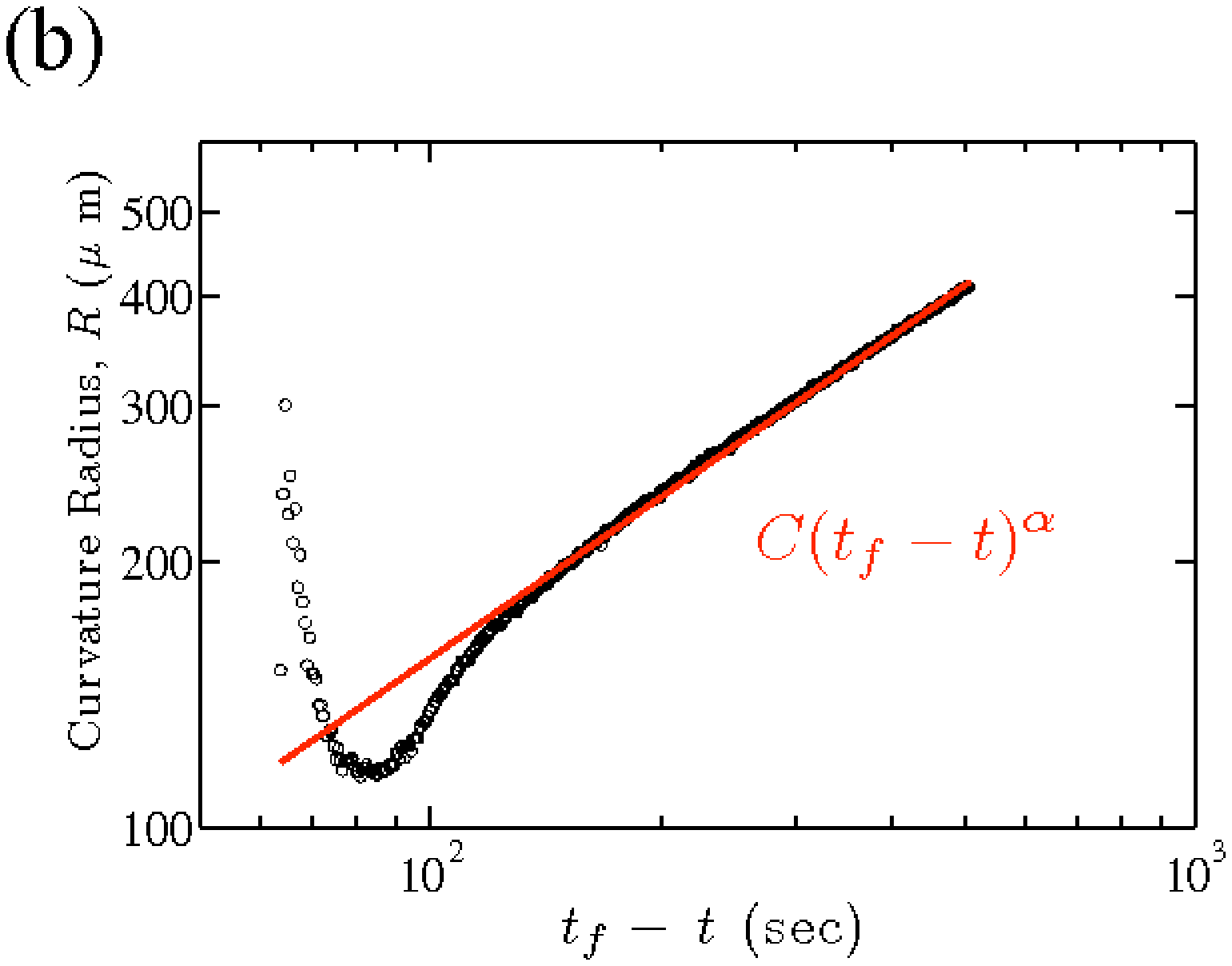}
\caption{\label{side-view-measurement} (Color online) (a) A typical contact angle dynamic of a water droplet evaporating on hydrophobic microstructure substrates of regular pillars of the width $w = 5 \unit{\mu m}$, the interspacing $a = 5 \unit{\mu m}$, and height $h = 6 \unit{\mu m}$ in a square lattice. The contact angle is calculated with the time evolutions of the drop height $H$ and the base radius $B$, shown in the inset, by assuming a spherical cap. The symbol $\ast$ marks the transition point from CB to W state. (b) The corresponding curvature radius $R$ reveals a power-law in time for $R \apprge 200~\unit{\mu m}$. The red line shows the best scaling fit of $C(t_f - t)^\alpha$, with the power-law exponent $\alpha = 0.61$.}\end{center}
\end{figure}

Complimentary to the bottom-views, side-view images as shown in Fig.~\ref{side-views} also provide useful information on the contact angle dynamics. Fig.~\ref{side-view-measurement}a shows the change of the macroscopic contact angle $\theta_{CA}$ in the course of the evaporation. $\theta_{CA}$ was calculated from the measurements of $H$ and $B$ by assuming a spherical water cap.  A kink after which a steep decrease in $\theta_{CA}$ can be noticed marks the wetting transition ($\ast$ in Fig.~\ref{side-view-measurement} a), while both $H$ and $B$ gradually decrease in time (see the inset of Fig.~\ref{side-view-measurement}a).  Fig.~\ref{side-view-measurement} b shows the curvature radius $R$ of the evaporating droplet. $R$ first decreases in time with a power-law relation and then increases again after the wetting transition. The red line is the best fit of the power-law $R(t) = C(t_f - t)^\alpha$, with the fitting results of $t_f = 505.7 \unit{s}$ and $\alpha = 0.61$. 
In comparison, for a pure diffusive evaporation, ignoring thermal and Marangoni convections, the evaporation rate is proportional to the perimeter of the droplet and the exponent then should be $1/2$~\cite{Deegan_Nature1997}.  Indeed, experiments with a variety of completely wetting liquids of alkanes drying on flat, isothermal, smooth, wetting surfaces found exponents close to $1/2$~\cite{Cachile_Langmuir_2002}, suggesting diffusive evaporation process for these completely wetting liquids. 
However, for an evaporating water droplet the exponent $\alpha \approx 0.60$ has been observed down to the drop radius $\approx 500~\mu m$ on a flat and perfectly wetting surface, such as mica, with different humidities and temperature~\cite{DBonn_JFM2006}. These observations might suggest a universal scaling for naturally evaporating water droplet with a freely moving contact line.  More importantly, in Fig.~\ref{side-view-measurement}b the data deviate from this power-law relation for droplet radii smaller than $200~\unit{\mu m}$ in the Wenzel transition state with the triple line pinned.

How to predict the critical drop size at the CB to W transition triggered by a natural evaporation?  Recent studies have proposed a ``touch down'' scenario for dilutely placed micro-pillars of $\Phi_s \approx 0.02$. That is, the transition happens when a dropping meniscus reaches the bottom of the sample between pillars by a local driving mechanism due to the Laplace excess pressure~\cite{Reyssat_EPL_2008, Jung_2008}.  However, for our densely packed micro-pillars the ``touch down'' scenario would predict the critical drop radius $R^\ast \sim a^2/h$ to be $\apprle 5 ~\unit{\mu m}$, which largely deviates from our experimental findings.  
For an alternative explanation of the observed critical size at the transition here we therefore estimate global surface energies $E_{CB}$ and $E_W$ for a CB and a Wenzel state, respectively.

The total surface energy $E_{CB}$ ($E_{W}$) is the sum of all energies needed for creating interfaces when placing a Fakir (Wenzel) drop onto the microstructures. During the evaporation, the drop size slowly changes with time and thus $E_{CB}$ and $E_W$ depend on size. The three interfacial tensions are denoted $\sigma_{ls}$, $\sigma_{sg}$, and $\sigma_{lg}$ for the liquid-solid, solid-gas, and liquid-gas interfaces, respectively.  Assuming flat menisci, $E_{CB} = N~[\sigma_{ls} w^2 + \sigma_{sg}(4wh+d^2-w^2)+\sigma_{lg}(d^2-w^2)] + \sigma_{lg} S_{cap}$, where $N$ is the number of asperities underneath the droplet base, and $S_{cap}$ is the surface area of the water cap entirely in contact with the air. Similarly, $E_W = N~[\sigma_{ls}(d^2+ 4wh)] + \sigma_{lg} S_{cap}$. According to our experimental condition, we used $\sigma_{lg} = 70~m\unit{N/m}$ for the surface tension of water and $\sigma_{sg} = 25~m\unit{N/m}$ for the PDMS surface~\cite{de_gennBook}. We then estimated the interfacial tension $\sigma_{ls}$ between water and PDMS surface by a force balance at the triple line. This is equivalent to the Young's relation: $\sigma_{lg}\cos{\theta_{CA}} = \sigma_{sg} - \sigma_{ls}$, where $\theta_{CA}$ is the macroscopic contact angle.  The change of contact angle during a natural evaporation has been measured numerously for hydrophilic~\cite{Cachile_Langmuir_2002, Erbil_Langmuir_2002,Guena_2006}, hydrophobic~\cite{Erbil_Langmuir_1999} (and the references therein) and, to a less extent, for superhydrophobic surfaces~\cite{MacHale_Langmuir_2005,Zhang_2006,Shin_2009}.  Yet for superhydrophobic samples very limited data have been reported in the literature and no rigorous theory has been developed for the contact angle evolutions during evaporation.  We used the experimental data of droplet height $H$ and base radius $B$ from the side-views to obtain the dependence of $\theta_{CA}$ on droplet size by assuming a spherical cap for the water droplet.

\begin{figure}
\begin{center}
\includegraphics[width=3in]{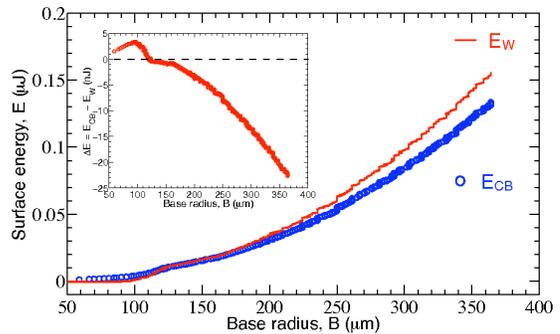}
\caption{\label{E_calc} (Color) Representative global interfacial energies of the Cassie-Baxster and Wenzel states: $E_{CB}$ and $E_W$, respectively, as a function of drop size. The inset shows the difference of these two energies, $\Delta E = E_{CB}- E_W$, giving the critical base radius $B^\ast$ for the wetting transition by the condition $\Delta E = 0$. Here, $B^{\ast} = 123~\mu m$.}
\end{center}
\end{figure}

Fig.~\ref{E_calc} shows a representative calculation of the total surface energy, depending on the drop size, for the CB and W state. When the base radius is large, $B \apprge 150 ~\unit{\mu m}$, the CB state is energetically favorable with a lower value of $E_{CB}$ than $E_W$.  The inset of Fig.~\ref{E_calc} shows the energy difference $\Delta E = E_{CB}-E_{W}$. For the wetting transition it holds $\Delta E(B^\ast) = 0$, giving the critical base radius $B^\ast$.  In this case, the predicted critical value of base radius is $B^\ast = 123~\unit{\mu m}$, which is in a good agreement with the experimental result $B_c = 125~\unit{\mu m}$, determined from the captured bottom images. 

\begin{figure}
\begin{center}
\includegraphics[width=3in]{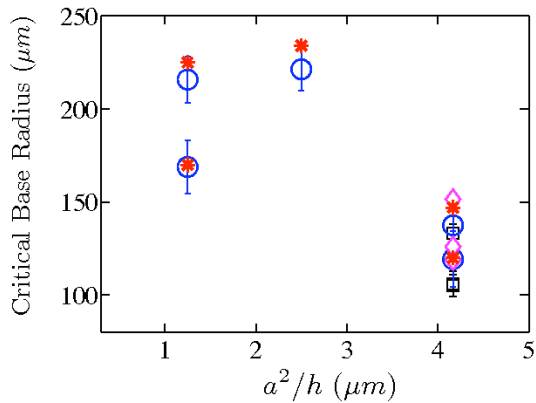}
\caption{\label{Bc} (Color online) The critical base radius at the evaporation triggered CB to W transition {\it vs.} the geometric parameter, $a^2/h$, for different initial droplet sizes.  Experimental data for $B_c$, marked by $\circ$ ($\Box$) for cylindrical (squared) micro-pillars in a squared lattice, agree well with the predicted $B^\ast$ shown by $\ast$ ($\diamond$) using the presented energy argument for the cylindrical (squared) micro-pillars.}
\end{center}
\end{figure}

We now extend our energy argument to various microstructures of different geometric arrangements. Fig.~\ref{Bc} shows the critical base radius for micropillars of different $h$ and pillar shapes (round/square), which can yield different surface roughness and packing fraction.  The energy estimate that predicts $B^\ast$ agrees very well with the critical size $B_c$ obtained from the bottom views. While the local ``touch down'' model predicts a linear increase of $B^\ast$ with the geometric parameter $a^2/h$, our data do not show this trend.  Instead, they agree well with our proposed interfacial energy model.

In summary, we experimentally monitored simultaneous side and bottom views of evaporating water droplets placed on hydrophobic micro-patterns.  The most essential observation is the change of the macroscopic contact angle during evaporation as the droplet gradually gets smaller and smaller.  At a certain size ($B^{\ast}$ a couple of $100~\mu m$) the initial water ``Fakir'' droplet jumps into the then energetically favorable Wenzel state. At the transition, the water infiltration dynamics starts at some nucleus and then propagates in a stepwise manner, which is profoundly affected by the geometric arrangements of the micro-pillars, in spite of the pinned triple line.  The successful predictions of the critical radius by the global interfacial energy argument is remarkable, as the transition from the CB to the W state is first only local.

\section{Acknowledgments} The authors gratefully thank Jacco Snoeijer for stimulating discussion and Alisia M. Peters for the micropatterned molds.


\end{document}